# Electrical Properties and Subband Occupancy at the (La,Sr)(Al,Ta)O$_3$/SrTiO$_3$ Interface

K. Han,[1,2] Z. Huang,[1,2,*] S. W. Zeng,[1,2] M. Yang,[3] C. J. Li,[4] W. X. Zhou,[1,2] X. Renshaw Wang,[5] T. Venkatesan,[1,2,4,6,7] J. M. D. Coey,[1,8] M. Goiran,[3] W. Escoffier,[3] and Ariando[1,2,6,*]

[1]*NUSNNI-NanoCore, National University of Singapore, Singapore 117411, Singapore.*

[2]*Department of Physics, National University of Singapore, Singapore 117542, Singapore.*

[3]*Laboratoire National des Champs Magnétiques Intenses (LNCMI-EMFL), CNRS-UGA-UPS-INSA, 143 Avenue de Rangueil, 31400 Toulouse, France*

[4]*Department of Materials Science and Engineering, National University of Singapore, Singapore 117575, Singapore.*

[5]*School of Physical and Mathematical Sciences & School of Electrical and Electronic Engineering, Nanyang Technological University, Singapore 639798, Singapore*

[6]*National University of Singapore Graduate School for Integrative Sciences and Engineering (NGS), 28 Medical Drive, Singapore 117456, Singapore.*

[7]*Department of Electrical and Computer Engineering, National University of Singapore, Singapore 117576, Singapore.*

[8]*School of Physics and CRANN, Trinity College, Dublin 2, Ireland*

*\*nnihz@nus.edu.sg, ariando@nus.edu.sg*

The quasi two-dimensional electron gas (q-2DEG) at oxide interfaces provides a platform for investigating quantum phenomena in strongly correlated electronic systems. Here, we study the transport properties at the high-mobility $(La_{0.3}Sr_{0.7})(Al_{0.65}Ta_{0.35})O_3$/$SrTiO_3$ (LSAT/STO) interface. Before oxygen annealing, the *as-grown* interface exhibits a high electron density and electron occupancy of two subbands: higher-mobility electrons ($\mu_1 \approx 10^4$ cm$^2$V$^{-1}$s$^{-1}$ at 2 K) occupy the lower-energy 3$d_{xy}$ subband, while lower-mobility electrons ($\mu_1 \approx 10^3$ cm$^2$V$^{-1}$s$^{-1}$ at 2 K) propagate in the higher-energy 3$d_{xz/yz}$-dominated subband. After removing oxygen vacancies by annealing in oxygen, only a single type of 3$d_{xy}$ electrons remain at the *annealed* interface, showing tunable Shubnikov-de Haas (SdH) oscillations below 9 T at 2 K and an effective mass of 0.7$m_e$. By contrast, no oscillation is observed at the *as-grown* interface even when electron mobility is increased to 50,000 cm$^2$V$^{-1}$s$^{-1}$ by gating voltage. Our results reveal the important roles of both carrier mobility and subband occupancy in tuning the quantum transport at oxide interfaces.

The SrTiO$_3$-based conducting interfaces such as LaAlO$_3$/SrTiO$_3$ (LAO/STO) have attracted much attention in the last decade. As the host of the quasi two-dimensional electron gas (q-2DEG) [1], the interface exhibits not only intriguing physics such as superconductivity and ferromagnetism that arises from strongly correlated electrons [2-11], but also attractive functionalities that are of potential for electronic devices [12-14]. Moreover, these STO-based interfaces with a q-2DEG provide an opportunity to study quantum transport in the strongly correlated 2D electronic system [15-21] that is different from the conventional semiconductor interface. However, the electron mobility of q-2DEG at STO-based interfaces is still relatively low. For example, the typical low-temperature electron mobility $\mu$ of conventional LAO/STO interfaces is only around 1000 cm$^2$V$^{-1}$s$^{-1}$ (or 0.1 m$^2$V$^{-1}$s$^{-1}$) [8,22-24]. To stimulate the Shubnikov-de Haas (SdH) oscillations, the magnetic field $B$ must exceed 10 T to fulfill the condition $\omega_c\tau = \mu B > 1$, where $\omega_c = Be/m^*$ is the cyclotron frequency, $e$ is the elementary charge, $m^*$ is the effective mass and $\tau$ is the transport elastic scattering time [16]. Such high magnetic fields make quantum transport measurement less accessible. In order to overcome this problem, many efforts have been made to improve the electron mobility at STO-based interfaces, such as modulating the surface charges/adsorbates [25], applying an external electrical field [26-28], and growing capping layers of $SrTiO_3/SrCuO_2$ [24] or a buffer layer of $(La,Sr)MnO_3$ [29]. Recent studies also showed that by replacing LAO with a different perovskite oxide $(La_{0.3}Sr_{0.7})(Al_{0.65}Ta_{0.35})O_3$ (LSAT), the electron mobility can easily reach $10^4$ $cm^2V^{-1}s^{-1}$ even after thermal annealing in oxygen [30,31]. This enhanced electron mobility can be explained by reducing the lattice mismatch [30,31] and polar discontinuity [32,33] at the interface. Hence, the high-mobility LSAT/STO interface is a good candidate for investigating and engineering low-field SdH oscillations (< 10 T) at the oxide interface.

Here, the transport properties of 20-unit-cell (uc) *as-grown* and *annealed* (001) LSAT/STO interfaces are compared. The samples were prepared by pulsed laser deposition (PLD) monitored by an *in-situ* reflection high-energy electron diffraction (RHEED) (see Fig. S1 in supplementary material [34]). For the *annealed* interface, the additional *ex-situ* annealing was performed at 600 $^o$C in 1 atm oxygen for 1 hour to remove oxygen vacancies produced during the PLD process [35]. Therefore, when compared to the *as-grown* interface, the *annealed* one should exhibit a lower electron density [35] and a better 2D confinement [36] of the q-2DEG, as shown in the inset of Fig. 1.

In Fig. 1(a), the temperature dependence of the sheet resistance $R_{xx}$ is shown for both *as-grown* and *annealed* LSAT/STO interfaces. Although $R_{XX}$ is higher at the *annealed* interface, both interfaces preserve the metallicity ($dR_{XX}/dT > 0$) down to 2 K. More interestingly, the Hall resistance $R_{XY}$ is quite different at these two interfaces. In Fig. 1(b), the *annealed* interface maintains the linear $R_{XY}$ from 300 to 2 K. On the other hand, the *as-grown* sample experiences a transition from the linear to nonlinear $R_{XY}$ below 50 K and the nonlinearity of $R_{XY}$ is enhanced by further cooling, as shown in Figure 1(c). Such nonlinear (linear) $R_{XY}$ is also found in other *as-grown* (*annealed*) LSAT/STO samples regardless of the LSAT thickness (see Fig. S2 in supplementary material [34]). Since the *as-grown* interface shows no sign of ferromagnetism or superparamagnetism at low temperatures (see Fig. S3 in supplementary material [34]), the nonlinear $R_{XY}$ is ascribed to the multiple types of carriers that have been reported in various STO-based heterostructures [27,37-42].

Considering there are two types of carriers coexisting at the interface, $R_{XY}$ and $R_{XX}$ can be written as

$$R_{\mathrm{XY}} = R_{\mathrm{H}}B = \frac{(\mu_1^2 n_1 + \mu_2^2 n_2) + (\mu_1 \mu_2 B)^2 (n_1 + n_2)}{e[(\mu_1|n_1| + \mu_2|n_2|)^2 + (\mu_1 \mu_2 B)^2 (n_1 + n_2)^2]} B \ , \quad (1)$$

and

$$R_{\mathrm{XX}} = \frac{1}{e(\mu_1|n_1| + \mu_2|n_2|)} \ , \quad (2)$$

where $R_H$ is the Hall coefficient, $n_1$, $n_2$ and $\mu_1$, $\mu_2$ are the carrier densities and mobilities for the two types of carriers [39-42]. As shown in Fig. 2(a), the experimental data (hollow symbols) of the nonlinear $R_{XY}$ can be nicely fitted by Eq. (1) (lines) with two types of electrons. Accordingly, the electron densities ($n_1$ and $n_2$) and mobilities ($\mu_1$ and $\mu_2$) used in the fitting are summarized in Fig. 2(b) and 2(c), respectively. Also, the electron density $n_{ann}$ and mobility $\mu_{ann}$ for the *annealed* interface are added for comparison. Clearly, these two types of electrons at the *as-grown* interface behave quite different. The first type of electrons exhibits the temperature-independent electron density $n_1$ and high electron mobility $\mu_1$ (= 11,000 cm$^2$V$^{-1}$s$^{-1}$ at 2 K). On the other hand, the second type of electrons shows the carrier freeze-out of $n_2$ and low electron mobility $\mu_2$ (= 1,000 cm$^2$V$^{-1}$s$^{-1}$ at 2 K). Given the similar temperature dependence of $n_1$ and $n_{ann}$ (as well as $\mu_1$ and $\mu_{ann}$), we argue that the high mobility (first type of) electrons are maintained at both the *as-grown* and *annealed* interfaces, while the low mobility (second type of) electrons are removed by the annealing.

Our previous study has shown that the oxygen vacancy at the *as-grown* interface can induce additional electrons [35], leading to a higher electron density compared to the *annealed* sample ($n_{total} = n_1+n_2 > n_{ann}$). On the other hand, the existence of oxygen vacancy could create the localized states slightly below the conduction band minimum of STO [35,43], resulting in the carrier freeze-out at low temperatures. By fitting $n_{total}$ of the *as-grown* LSAT/STO interface as a function of temperature (see Fig. S4 in supplementary material [34]), the estimated energy gap between oxygen vacancies and conduction band is around 5.6 meV, which is close to the value obtained at the *as-grown* LAO/STO interface (~ 4.2 meV) [35]. Furthermore, it should be noticed that only $n_2$ exhibits the carrier freeze-out, while $n_1$ is almost unchanged on cooling. This means that the first type (second type) of electrons occupies the low-energy (high-energy) state in the conduction band. It also explains why the first type of high-mobility electrons can survive at the *annealed* interface with a lower Fermi level.

To better understand these two types of electrons, transport experiments were conducted on the *as-grown* sample with applying a back-gating electric potential $V_g$. The inset of Fig. 3(a) shows a sketch of the back-gating setup. Before recording the experimental results, a field scan from -200 to 200 V is performed to minimize the hysteresis and/or artificial effect from trapping centers (see Fig. S5 in supplementary material [34]). In Fig. 3(a) of Hall effect, the nonlinearity of $R_{XY}$ is gradually suppressed by applying the negative $V_g$, and vice versa. By using Eq. (1) to fit these $R_{XY}$ curves with different $V_g$, the corresponding $n_1$ and $n_2$, as well as $\mu_1$ and $\mu_2$, are displayed in Fig. 3(b) and 3(c), respectively. For carrier densities, both $n_1$ and $n_2$ can be effectively tuned by $V_g$, but obviously $n_2$ is more sensitive to the back-gating electric field. Carrier mobility, $\mu_1$ increases from 2,000 to 50,000 cm$^2$V$^{-1}$s$^{-1}$ when changing $V_g$ from -200 to 200 V; while $\mu_2$ is around 700-900 cm$^2$V$^{-1}$s$^{-1}$ and independent of $V_g$. A similar $V_g$ dependence of electron densities and mobilities was also reported at the LAO/STO interface [16,26].

It has been well documented that the symmetry breaking at the STO-based interface can lift the degeneracy of Ti 3$d$ orbitals and result in a band structure with multiple conducting subbands [39,44-52]. In particular, it is a widespread view that the lower-energy subband is formed from the Ti 3$d_{xy}$ orbitals, which are characterized by the strong 2D confinement and a small effective mass $m^* \sim 0.7 m_e$ in the $xy$ plane ($m_e$ is the free electron mass); whereas the higher-energy subband is dominated by Ti 3$d_{xz/yz}$ orbitals, the electrons of which have a larger $m^* \sim 2\text{-}3 m_e$ and propagate deeper into STO substrate [45-52]. At the *as-grown* LSAT/STO interface, oxygen vacancies increase the carrier density and raise the position of Fermi level $E_F$. Therefore, the electrons may occupy both subbands – the first type of electrons ($n_1$ and $\mu_1$) are in the lower-energy/lighter 3$d_{xy}$ subband and the second type ($n_2$ and $\mu_2$) are in the higher-energy/heavier 3$d_{xz/yz}$-dominated subband. For the *annealed* interface, both the carrier density and Fermi level are lowered because oxygen vacancies have been removed, and only the $3d_{xy}$ subband is populated. This picture can nicely explain not only the different $R_{XY}$ at the *as-grown* and *annealed* interfaces, but also the $V_g$-dependent transport behavior. Because the density of states (DOS) of $3d_{xz/yz}$ electrons is much larger than that of the $3d_{xy}$ electrons near the Fermi level [39,45,46,51], $n_2$ can be more effectively tuned by $V_g$ than $n_1$, as shown in Fig. 3(b). The mobility $\mu_1$ for the electrons at the interface must be strongly influenced by the potential fluctuations arising from interfacial disorder [47,48,53], while $\mu_2$ for the electrons deeper in the STO bulk is limited by the large effective mass of $3d_{xz/yz}$ orbitals in the *xy* plane [45-52]. When applying a positive $V_g$, the increased band bending moves electrons away from the interface, which will reduce the effect of potential fluctuation but it has less influence on the effective mass. Thus, $\mu_1$ can always benefit from the positive $V_g$ but it will have little effects on $\mu_2$ as seen in Fig. 3(c).

Our transport data have shown that the high-mobility electrons (~ 11,000 $cm^2V^{-1}s^{-1}$ at 2 K) are present at both *as-grown* and *annealed* LSAT/STO interfaces. Therefore, both interfaces are expected to show low-field SdH oscillations. This is not the case. For the *as-grown* interface, Figure 4(a) and 4(b) display the derivative of the magnetoresistance ($dR_{xx}/dB$) and the corresponding $\Delta R_{xx}$ ($\Delta R_{xx} = R_{xx} - R_B$, where $R_B$ is the fitted polynomial background) with different $V_g$ at 2 K. Surprisingly, there are no observable oscillations even at $V_g$ = 200 V when $\mu_1$ raises to 50,000 $cm^2V^{-1}s^{-1}$. In contrast, the *annealed* interface shows clear SdH oscillations in both $dR_{xx}/dB$ and $\Delta R_{xx}$ in Fig. 4(c) and 4(d). Moreover, these oscillations are enhanced by applying a larger positive $V_g$ with a higher electron mobility. Figure 4(e) displays the Fourier transform (FT) of the oscillating $\Delta R_{xx}$ from the *annealed* interface under 120 V. A single FT frequency peak is observed at 39.8 T, corresponding to the SdH periodicity $1/\Delta B \sim 0.025$ T$^{-1}$. Thus, the 2D electron density in a circular Fermi surface can be estimated by $n_{2D} = \Delta B g_S g_V e/h$, where $g_S$ and $g_V$ are the spin and valley degeneracy, respectively. If we assume $g_S = 2$ and $g_V = 1$, $n_{2D}$ can be calculated as $1.92 \times 10^{12}$ cm$^{-2}$, which is only one ninth of $n_{ann} \sim 1.75 \times 10^{13}$ cm$^{-2}$ (see Fig. S6 in supplementary material [34]). Previous studies found that the value of $n_{2D}$ is about one sixth of the carrier densities measured from Hall effect at the LAO/STO interface [15,16]. For SdH oscillations, the oscillating amplitude $\Delta R_{xx}$ can be written as

$$\Delta R_{XX}(T) = 4R_0 e^{-\alpha T_D} \frac{\alpha T}{\sinh(\alpha T)} \quad (3)$$

where $R_0$ is the non-oscillatory component of $R_{XX}$, $\alpha$ is estimated by $\alpha = 2\pi^2 k_B/\hbar \omega_C$, $\omega_C = Be/m^*$, and $T_D$ is the Dingle temperature [15]. By fitting the temperature-dependent $\Delta R_{xx}/R_0$ in Fig. 4(f), the high-mobility (first type of) electron is characterized by $m^* = 0.70 \pm 0.04\, m_0$ and $T_D = 4.4 \pm 0.3$ K. The small effective mass ($m^* < m_0$) supports our statement that the higher-mobility electrons are travelling in the $3d_{xy}$ orbital.

We argue that the discrepancy between $n_{2D}$ and $n_{ann}$ at the STO-based interface arises from a simplification of the subband structure. It should be noted that the degree of interfacial symmetry breaking is different for each STO layer where the q-2DEG resides [45-50]. When the STO layer is located closer to the interface, the symmetry breaking is stronger, resulting in a larger splitting energy between the $3d_{xy}$ and $3d_{xz/yz}$ orbitals. To sum up the non-degenerate $3d_{xy}$ orbitals from all STO layers, there must be several $3d_{xy}$ channels with the same low effective mass but different energy. In this case, $g_V$ will be larger than 1, which may explain the discrepancy between $n_{2D}$ and $n_{ann}$ observed at our LSAT/STO and LAO/STO interfaces [15,16]. Our argument can be supported by recent quantum Hall measurements, which have shown that there are 4 to 10 channels at the SrTiO$_3$-based interfaces [18,21].

Recent studies have suggested that the SdH oscillations in high fields arise from electrons in the 3$d_{xz/yz}$ subband at the LAO/STO interface [17-20], while the 3$d_{xy}$ electrons are localized with small electron mobility below a mobility edge (ME) created by interfacial disorder [47,48,53]. The lattice mismatch of LAO/STO is 3%, while it is only 1% for LSAT/STO. Correspondingly, the position of the mobility edge should be lower at the LSAT/STO interface as shown in Figure 4(g). Hence, some of the 3$d_{xy}$ electrons, which are less mobile at the LAO/STO interface, could recover their high mobility with $m^* \sim 0.7 m_e$ and low-field SdH oscillations at the LSAT/STO interface.

The absence of SdH oscillations at the *as-grown* interface can be understood in terms of the position of Fermi level in the two-subband band structure. As shown in Fig. 4(g), the extra oxygen-vacancy-induced mobile electrons move the Fermi level of the *as-grown* interface to the high energy states which are dominated by the heavy 3$d_{xz/yz}$ subband [39,45-52]. When a magnetic field of around 9 T is applied, the DOS of the 3$d_{xz/yz}$-dominated subband will hardly form separated Landau levels below 9 T, because the broadening of Landau levels $k_B T_D$ is comparable to the energy separation $\hbar \omega_C$ ($\hbar \omega_C$ = 0.5 meV with $m^* = 2 m_e$ and $B$ = 9 T; $k_B T_D$ = 0.4 meV with $T_D$ = 4.4 K). In this case, the 3$d_{xy/xz}$ oscillations could be very weak and hard to measure. Furthermore, 3$d_{xy}$ electrons above the band crossing (Lifshitz point) strongly interact with 3$d_{xz/yz}$ orbitals due to the enhanced spin-orbital interactions in the high energy 3$d_{xz/yz}$-dominated subband [39,46]. Hence, the mobile electrons above the band crossing lose their two dimensional character with a variable DOS [39,46] and give no observable oscillations at low field even when their mobility is very high. On the other hand, Figure 4(h) shows that with lower carrier density the Fermi level of the *annealed* interface lies within the 2D low-energy $3d_{xy}$ subband with a constant DOS. Under a 9 T magnetic field, the high electron mobility and small effective mass can then ensure separated Laudau levels with $\mu_1 B \sim 1$ m$^2$V$^{-1}$s$^{-1}$ × 9 T > 1 and $\hbar\omega_C > k_B T_D$ ($\hbar\omega_C$ = 1.5 meV with $m^* = 0.7 m_e$), explaining why only the *annealed* interface exhibits SdH oscillations below 9 T. Our data reveal that pushing the electron mobility alone may not guarantee the low-field SdH oscillations at the STO-based interfaces, while the subband occupancy also plays a crucial role. Additionally, $V_g$-dependent transport data demonstrate that the nonlinearity of $R_{XY}$ for the *as-grown* interface can be preserved with $n_1 + n_2 \geq 1.94 \times 10^{13}$ cm$^{-2}$, and the linearity of $R_{XY}$ for the *annealed* interface with $n_{ann} \leq 1.78 \times 10^{13}$ cm$^{-2}$. It suggests that the mobile electrons at the LSAT/STO interface begin to occupy the $3d_{xz/yz}$-dominated subband when carrier density is within the range of 1.78–1.96×10$^{13}$ cm$^{-2}$, which is close to the universal carrier density for a Lifshitz transition at LAO/STO interfaces of 1.50–1.86 ×10$^{13}$ cm$^{-2}$ [39].

In summary, we have compared the transport properties of the *as-grown* and *annealed* LSAT/STO interfaces. The *as-grown* interface is characterized by a higher carrier density and nonlinear Hall effect, which is explained by the coexistence of two types of electrons. Detailed studies of the magnetic- and electric-field-dependent transport data suggested that the higher-mobility electrons ($\mu_1 \sim$ 11,000 cm$^2$V$^{-1}$s$^{-1}$ at 2 K) are located close to the interface and occupy the low-energy $3d_{xy}$ subband; while the lower-mobility electrons ($\mu_2 \sim$ 1,000 cm$^2$V$^{-1}$s$^{-1}$ at 2 K) travel deeper in the STO substrate in the higher-energy $3d_{xz/yz}$-dominated subband. After removing the oxygen vacancies, only the low-energy/high-mobility $3d_{xy}$ electrons remain at the *annealed* interface, showing tunable SdH oscillations below 9 T at 2 K and an effective mass ~ 0.7$m_e$. Our results highlight the effect of subband occupancy in manipulating the low-field quantum transport at oxide interfaces.

This work is supported by the grant NEXT n° ANR-10-LABX-0037 in the framework of the "Programme des Investissements d'Avenir", MOE Tier 1 (Grants No. R-144-000-364-112 and No. R-144-000-346-112) and Singapore National Research Foundation (NRF) under the Competitive Research Programs (CRP Awards No. NRF-CRP8-2011-06, No. NRF-CRP10-2012-02, and No. NRF-CRP15-2015-01).

1. A. Ohtomo and H. Y. Hwang, Nature (London) **427**, 423 (2004).

2. N. Reyren, S. Thiel, A. D. Caviglia, L. Fitting Kourkoutis, G. Hammerl, C. Richter, C. W. Schneider, T. Kopp, A.-S. Rüetschi, D. Jaccard, M. Gabay, D. A. Muller, J.-M. Triscone, and J. Mannhart, Science **317**, 1196 (2007).

3. C. Richter, H. Boschker, W. Dietsche, E. Fillis-Tsirakis, R. Jany, F. Loder, L. F. Kourkoutis, D. A. Muller, J. R. Kirtley, C. W. Schneider, and J. Mannhart, Nature (London) **502**, 528 (2013).

4. A. Brinkman, M. Huijben, M. van Zalk, J. Huijben, U. Zeitler, J. C. Maan, W. G. van der Wiel, G. Rijnders, D. H. A. Blank, and H. Hilgenkamp, Nat. Mater. **6**, 493 (2007).

5. F. Bi, M. Huang, S. Ryu, H. Lee, C. W. Bark, C. B. Eom, P. Irvin, and J. Levy, Nat. Commun. **5**, 5019 (2014).

6. A. D. Caviglia, M. Gabay, S. Gariglio, N. Reyren, C. Cancellieri, and J.-M. Triscone, Phys. Rev. Lett. **104**, 126803 (2010).

7. M. Ben Shalom, M. Sachs, D. Rakhmilevitch, A. Palevski, and Y. Dagan, Phys. Rev. Lett. **104**, 126802 (2010).

8. Ariando, X. Wang, G. Baskaran, Z. Q. Liu, J. Huijben, J. B. Yi, A. Annadi, A. Roy Barman, A. Rusydi, S. Dhar, Y. P. Feng, J. Ding, H. Hilgenkamp, and T. Venkatesan, Nat. Commun. **2**, 188 (2011).

9. L. Li, C. Richter, J. Mannhart, and R. C. Ashoori, Nat. Phys. **7**, 762 (2011).

10. D. A. Dikin, M. Mehta, C. W. Bark, C. M. Folkman, C. B. Eom, and V. Chandrasekhar, Phys. Rev. Lett. **107**, 056802 (2011).

11. J. A. Bert, B. Kalisky, C. Bell, M. Kim, Y. Hikita, H. Y. Hwang, and K. A. Moler, Nat. Phys. **7**,

767 (2011).

12. C. Cen, S. Thiel, J. Mannhart, and J. Levy, Science **323**, 1026 (2009).

13. T. D. N. Ngo, J.-W. Chang, K. Lee, S. Han, J. S. Lee, Y. H. Kim, M.-H. Jung, Y.-J. Doh, M.-S. Choi, J. Song, and J. Kim, Nat. Commun. **6**, 8035 (2015).

14. B. Förg, C. Richter and J. Mannhart, Appl. Phys. Lett. **100**, 053506 (2012).

15. A. D. Caviglia, S. Gariglio, C. Cancellieri, B. Sacépé, A. Fête, N. Reyren, M. Gabay, A. F. Morpurgo, and J.-M. Triscone, Phys. Rev. Lett. **105**, 236802 (2010).

16. M. Ben Shalom, A. Ron, A. Palevski, and Y. Dagan, Phys. Rev. Lett. **105**, 206401 (2010).

17. A. McCollam, S. Wenderich, M. K. Kruize, V. K. Guduru, H. J. A. Molegraaf, M. Huijben, G. Koster, D. H. A. Blank, G. Rijnders, A. Brinkman, H. Hilgenkamp, U. Zeitler, and J. C. Maan, APL Mater. **2**, 022102 (2014).

18. Y. W. Xie, C. Bell, M. Kim, H. Inoue, Y. Hikita, and H. Y. Hwang, Solid State Commun. **197**, 25 (2014).

19. A. Fête, S. Gariglio, C. Berthod, D. Li, D. Stornaiuolo, M. Gabay, and J.-M. Triscone, New J. Phys. **16**, 112002 (2014).

20. M. Yang, K. Han, O. Toressin, M. Pierre, S. W. Zeng, Z. Huang, T. V. Venkatesan, M. Goiran, J. M. D. Coey, Ariando, and W. Escoffier, Appl. Phys. Lett. **109**, 122106 (2016).

21. F. Trier, G. E. D. K. Prawiroatmodjo, Z. C. Zhong, D. V. Christensen, M. von Soosten, A. Bhowmik, J. M. G. Lastra, Y. Z. Chen, T. S. Jespersen, and N. Pryds, Phys. Rev. Lett. **117**, 096804 (2016).

22. J. Mannhart and D. G. Schlom, Science **327**, 1607 (2010).

23. G. Herranz, F. Sánchez, N. Dix, M. Scigaj, and J. Fontcuberta, Sci. Rep. **2**, 758 (2012).

24. M. Huijben, G. Koster, M. K. Kruize, S. Wenderich, J. Verbeeck, S. Bals, E. Slooten, B. Shi, H. J. A. Molegraaf, J. E. Kleibeuker, S. van Aert, J. B. Goedkoop, A. Brinkman, D. H. A. Blank, M. S. Golden, G. van Tendeloo, H. Hilgenkamp, and G. Rijnders, Adv. Funct. Mater. **23**, 5240 (2013).

25. Y. W. Xie, C. Bell, Y. Hikita, S. Harashima, and H. Y. Hwang, Adv. Mater. **25**, 4735 (2013).

26. S. W. Zeng, W. M. Lü, Z. Huang, Z. Q. Liu, K. Han, K. Gopinadhan, C. J. Li, R. Guo, W. X. Zhou, H. J. Harsan Ma, L. K. Jian, T. Venkatesan, and Ariando, ACS Nano **10**, 4532 (2016).

27. C. Bell, S. Harashima, Y. Kozuka, M. Kim, B. G. Kim, Y. Hikita, and H. Y. Hwang, Phys. Rev. Lett. **103**, 226802 (2009).

28. M. Hosoda, Y. Hikita, H. Y. Hwang, and C. Bell, Appl. Phys. Lett. **103**, 103507(2013).

29. Y. Z. Chen, F. Trier, T. Wijnands, R. J. Green, N. Gauquelin, R. Egoavil, D. V. Christensen G. Koster, M. Huijben, N. Bovet, S. Macke, F. He, R. Sutarto, N. H. Andersen, J. A. Sulpizio, M. Honig, G. E. D. K. Prawiroatmodjo, T. S. Jespersen, S. Linderoth, S. Ilani, J. Verbeeck, G. Van Tendeloo, G. Rijnders, G. A. Sawatzky, and N. Pryds, Nat. Mater. **14**, 801 (2015).

30. Z. Huang, K. Han, S. W. Zeng, M. Motapothula, A. Y. Borisevich, S. Ghosh, W. M. Lü, C. J. Li, W. X. Zhou, Z. Q. Liu, M. Coey, T. Venkatesan , and Ariando Nano Lett. **16**, 2307 (2016).

31. K. Han, N. Palina, S. W. Zeng, Z. Huang, C. J. Li, W. X. Zhou, D. Y. Wan, L. C. Zhang, X. Chi, R. Guo, J. S. Chen, T. Venkatesan, A. Rusydi, and Ariando, Sci. Rep. **6**, 25455 (2016).

32. F. Gunkel, R. Waser, A. H. H. Ramadan, R. A. De Souza, S. Hoffmann-Eifert, and R. Dittmann, Phys. Rev. B **93**, 245431 (2016).

33. F. Gunkel, S. Hoffmann-Eifert, R. A. Heinen, D. V. Christensen, Y. Z. Chen, R. Waser, and R.

Dittmann, ACS Appl. Mater. Interfaces **9**, 1086 (2017).

34. See Supplementary Material at xxx for a detailed description of sample preparation, nonlinear Hall effect and magnetic measurement of the *as-grown* sample, estimation of the oxygen-vacancy activation energy, and electric gating experiments for the *as-grown* and *annealed* interface.

35. Z. Q. Liu, C. J. Li, W. M. Lü, X. H. Huang, Z. Huang, S. W. Zeng, X. P. Qiu, L. S. Huang, A. Annadi, J. S. Chen, J. M. D. Coey, T. Venkatesan, and Ariando, Phys. Rev. X **3**, 021010 (2013).

36. M. Basletic, J.-L. Maurice, C. Carrétéro, G. Herranz, O. Copie, M. Bibes, É. Jacquet, K. Bouzehouane, S. Fusil, and A. Barthélémy, Nat. Mater. **7**, 621 (2008).

37. A. Fête, C. Cancellieri, D. Li, D. Stornaiuolo, A. D. Caviglia, S. Gariglio and J.-M. Triscone Appl. Phys. Lett. **106**, 051604 (2015).

38. J. Biscaras, N. Bergeal, S. Hurand, C. Grossetête, A. Rastogi, R. C. Budhani, D. LeBoeuf, C. Proust, and J. Lesueur, Phys. Rev. Lett. **108**, 247004 (2012).

39. A. Joshua, S. Pecker, J. Ruhman, E. Altman, and S. Ilani, Nat. Commun. **3**, 1129 (2012).

40. J. S. Kim, S. S. A. Seo, M. F. Chisholm, R. K. Kremer, H.-U. Habermeier, B. Keimer, and H. N. Lee, Phys. Rev. B **82**, 201407(R) (2010).

41. R. Ohtsuka, M. Matvejeff, K. Nishio, R. Takahashi, and M. Lippmaa, Appl. Phys. Lett. **96**, 192111 (2010).

42. X. Renshaw Wang, L. Sun, Z. Huang, W. M. Lü, M. Motapothula, A. Annadi, Z. Q. Liu, S. W. Zeng, T. Venkatesan & Ariando, Sci. Rep. **5**, 18282 (2015).

43. Z. Q. Liu, D. P. Leusink, X. Wang, W. M. Lü, K. Gopinadhan, A. Annadi, Y. L. Zhao, X. H.

Huang, S. W. Zeng, Z. Huang, A. Srivastava, S. Dhar, T. Venkatesan, and Ariando, Phys. Rev. Lett. **107**, 146802 (2011).

44. M. Salluzzo, J. C. Cezar, N. B. Brookes, V. Bisogni, G. M. De Luca, C. Richter, S. Thiel, J. Mannhart, M. Huijben, A. Brinkman, G. Rijnders, and G. Ghiringhelli, Phys. Rev. Lett. **102**, 166804 (2009).

45. J. R. Tolsma, A. Principi, R. Asgari, M. Polini, and A. H. MacDonald, Phys. Rev. B **93**, 045120 (2016).

46. N. Mohanta and A. Taraphder, Phys. Rev. B **92**, 174531 (2015).

47. P. Delugas, A. Filippetti, V. Fiorentini, D. I. Bilc, D. Fontaine, and P. Ghosez, Phys. Rev. Lett. **106**, 166807 (2011).

48. Z. S. Popović, S. Satpathy, and R. M. Martin, Phys. Rev. Lett. **101**, 256801 (2008).

49. W.-J. Son, E. Cho, B. Lee, J. Lee, and S. Han, Phys. Rev. B **79**, 245411 (2009).

50. M. Stengel, Phys. Rev. Lett. **106**, 136803 (2011).

51. A. P. Petrović, A. Paré, T. R. Paudel, K. Lee, S. Holmes, C. H. W. Barnes, A. David, T. Wu, E. Y. Tsymbal, and C. Panagopoulos, Sci. Rep. **4**, 5338 (2014).

52. L. W. van Heeringen, G. A. de Wijs, A. McCollam, J. C. Maan, and A. Fasolino, Phys. Rev. B **88**, 205140 (2013).

53. Z. Huang, X. Renshaw Wang, Z. Q. Liu, W. M. Lü, S. W. Zeng, A. Annadi, W. L. Tan, X. P. Qiu, Y. L. Zhao, M. Salluzzo, J. M. D. Coey, T. Venkatesan, and Ariando, Phys. Rev. B **88**, 161107(R) (2013).

**Figure 1** (a) The temperature-dependent sheet resistance $R_{XX}$ for the 20 uc *as-grown* and *annealed* LSAT/STO interface. Inset: the sketch of q-2DEG without (with) oxygen vacancy at the *annealed* (*as-grown*) interface. (b) The linear Hall effect of the *annealed* interface. (c) The nonlinear Hall effect of the *as-grown* interface below 100 K.

**Figure 2** (a) The nonlinear Hall effect of *as-grown* interface at different temperatures. The offset of $R_{XY}$ is set for clarity. The experimental data are denoted by open symbols, while the fitting data by solid lines. (b) The temperature dependence of carrier densities. (c) The temperature dependence of electron mobilities. Two types of electrons at the *as-grown* interface are denoted by $n_1$, $n_2$, $\mu_1$ and $\mu_2$, while the single type at the *annealed* interface is $n_{ann}$ and $\mu_{ann}$.

**Figure 3** (a) $R_{XY}$ of the *as-grown* interface at 2 K under various $V_g$. Inset: the sketch for the back-gating setup. (b) Two carrier densities as a function of $V_g$. Inset: the sketch of the subband structure considering the spin-orbital interactions [39,46]. Below the band crossing, the low-energy subband is formed by pure $3d_{xy}$ orbitals (red curves). Above the band crossing, the strong spin-orbital interactions result in mixing of the $3d_{xy}$ (red curves) and $3d_{xz/yz}$ orbitals (blue curves) in the high-energy $3d_{xz/yz}$-dominated subband. (c) Two carrier mobilities as a function of $V_g$. The dash line in (a) and solid lines in (b) and (c) are plotted to guide the eye.

**Figure 4** (a) The d$R_{xx}$/d$B$ and (b) the corresponding Δ$R_{xx}$ of the *as-grown* interface under various $V_g$. (c) The d$R_{xx}$/d$B$ and (d) the corresponding Δ$R_{xx}$ of the *annealed* interface under various $V_g$. The offset of Δ$R_{xx}$ is set for clarity. (e) Fourier spectrum of Δ$R_{xx}$ measured at 2 K/120 V from *annealed* interface. (f) Orange dots: temperature dependence of Δ$R_{xx}/R_0$ at 7.8 T/0 V from *annealed* interface. (g) DOS of the *as-grown*/high-density interface before and after applying the magnetic field. The lines of ME denote the position of mobility edge at the LAO/STO and LSAT/STO interfaces, respectively. (h) DOS of the *annealed*/low-density interface before and after applying the magnetic field.

Figure 1

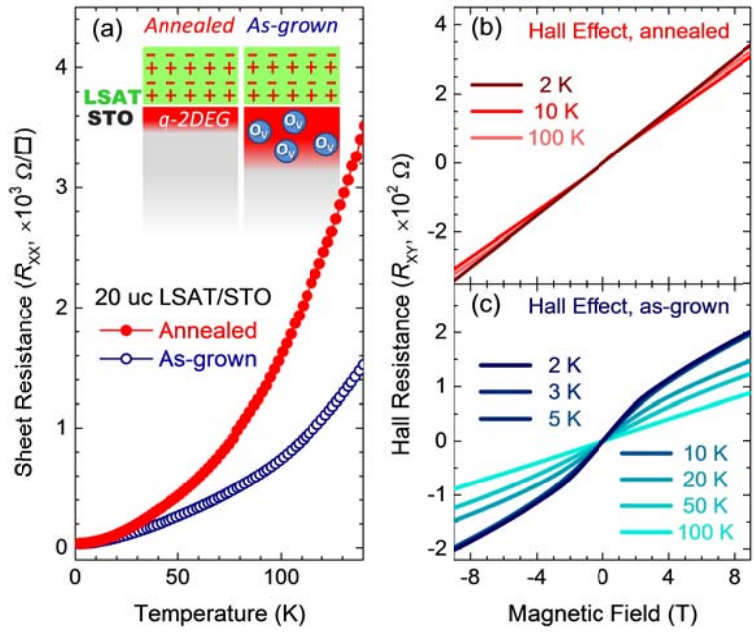

Figure 2

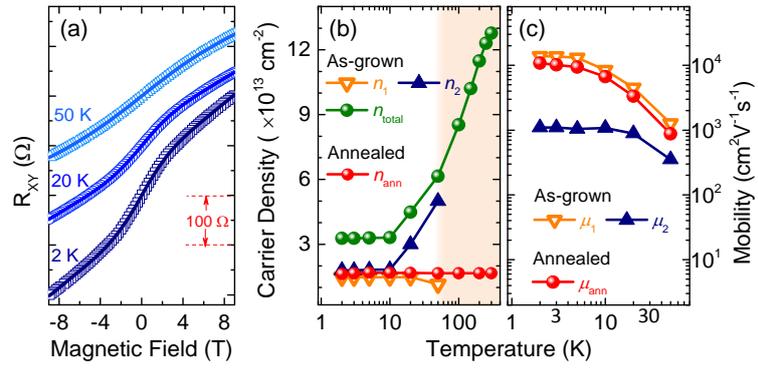

Figure 3

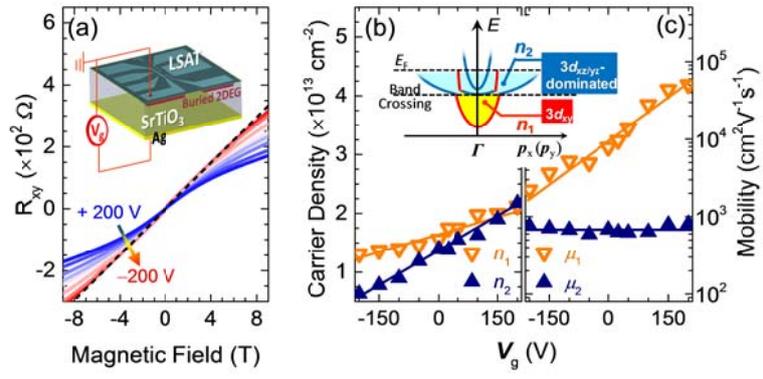

Figure 4

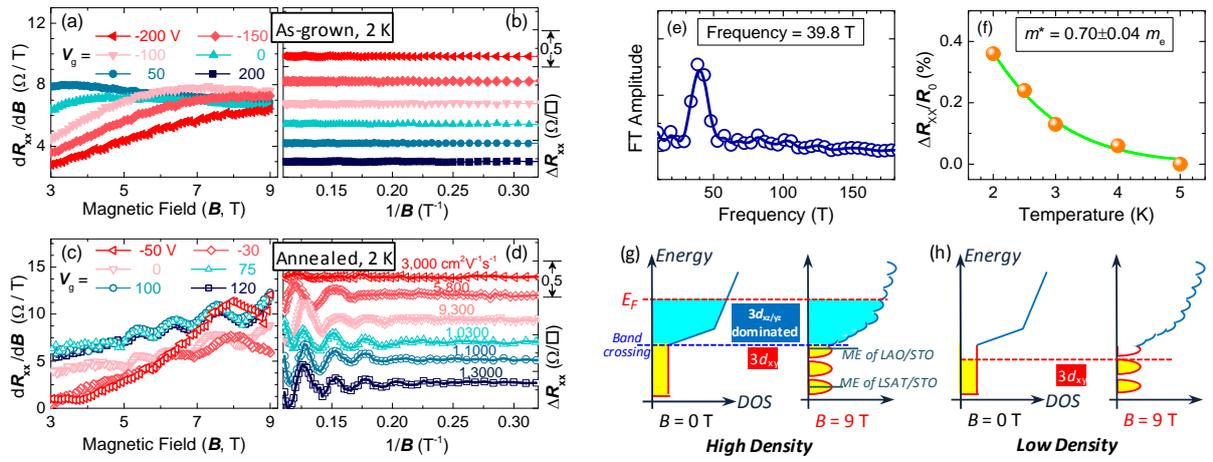